\documentclass[prc,twocolumn,showpacs,preprintnumbers]{revtex4}
\usepackage{bm}
\usepackage{amssymb}
\usepackage{amsmath}
\usepackage{graphicx}

\begin{document}
\title{Diffusion of Neon in White Dwarf Stars}
\author{J. Hughto}\email{jhughto@indiana.edu}
\author{A. S. Schneider}
\author{C. J. Horowitz}\email{horowit@indiana.edu} 
\affiliation{Department of Physics and Nuclear Theory Center,
             Indiana University, Bloomington, IN 47405}
\author{D. K. Berry}
\affiliation{University Information Technology Services,
             Indiana University, Bloomington, IN 47408}

\date{\today}
\begin{abstract}
Sedimentation of the neutron rich isotope $^{22}$Ne may be an important source of gravitational energy during the cooling of white dwarf stars.  This depends on the diffusion constant for $^{22}$Ne in strongly coupled plasma mixtures.  We calculate self-diffusion constants $D_i$ from molecular dynamics simulations of carbon, oxygen, and neon mixtures.  We find that $D_i$ in a mixture does not differ greatly from earlier one component plasma results.  For strong coupling (coulomb parameter $\Gamma>$ few), $D_i$ has a modest dependence on the charge $Z_i$ of the ion species, $D_i \propto Z_i^{-2/3}$.  However $D_i$ depends more strongly on $Z_i$ for weak coupling (smaller $\Gamma$).   We conclude that the self-diffusion constant $D_{\rm Ne}$ for $^{22}$Ne in carbon, oxygen, and neon plasma mixtures is accurately known so that uncertainties in $D_{\rm Ne}$ should be unimportant for simulations of white dwarf cooling.

\end{abstract}
\smallskip


\maketitle

\section{Introduction}
\label{sec.introduction}

Observations of cooling white dwarf (WD) stars provide important ways to date stellar systems \cite{cosmochron}.   White dwarf cooling involves simple physics because the energy from nuclear reactions is modest.  However, chemical energy from the latent heat of fusion and gravitational energy from sedimentation can be important.  Understanding these additional energy sources may allow more accurate stellar dates, see for example \cite{nature465_2010_194}, and provide additional information on the interior composition of WDs \cite{WD_PRL}.

The interior of a WD is a coulomb plasma of ions and a degenerate electron gas.  As the star cools this plasma crystallizes.  Winget et al. recently observed effects from the latent heat of crystallization on the luminosity function of WDs in the globular cluster NGC 6397 \cite{winget}.   Winget et al.'s observations constrain the melting temperature of the carbon and oxygen mixtures expected in these WD cores.  This temperature depends on the ratio of carbon to oxygen.    Therefore observations of crystallization may provide information on WD composition.

The ratio of carbon to oxygen in WD stars is very interesting.  It depends on the rate for the reaction $^{12}$C($\alpha,\gamma$)$^{16}$O.  Despite a great deal of effort, see for example \cite{c12ag}, the stellar rate for this reaction remains one of the most important unsettled rates left in Nuclear Astrophysics \cite{buchmann}.  Furthermore, the ratio of carbon to oxygen in massive stars is important for their subsequent evolution and nucleosynthesis \cite{coevolution}.   Therefore, a measurement of the carbon to oxygen ratio in a WD could be very important.

To determine the C/O ratio from observations of the melting temperature, one needs the phase diagram for carbon and oxygen mixtures.  Recently, we determined this phase diagram using molecular dynamics simulations and concluded that, if the melting temperature is close to that for pure carbon, then the central oxygen fraction by mass of the carbon/oxygen WDs in NGC 6397 is less than about 64 \% \cite{WD_PRL}.  Note that the interior of WDs can also be probed with astro-seismology, see for example ref. \cite{Metcalfe}. 

Sedimentation of $^{22}$Ne could provide an additional energy source during WD cooling \cite{Ne_lars,Ne_lars2}.  Much of the carbon, nitrogen, and oxygen, originally present in the star, is converted by nuclear reactions into $^{22}$Ne.  This neutron rich isotope, with 12 neutrons and 10 protons, has a larger mass to charge ratio than $^{12}$C or $^{16}$O.  As a result it will sink in the strong gravitational field of the star and release gravitational energy.  Recently, Garcia-Berro et al. \cite{apj677_2008_473} studied the effects of $^{22}$Ne on WD evolution.  Sedimentation could be most important in very metal rich stars that have more $^{22}$Ne.  Both the release of latent heat from crystallization and gravitational energy from sedimentation can delay WD cooling.  Garcia-Berro et al. \cite{nature465_2010_194} can explain the long 8 Gyr age for the metal rich open star cluster NGC 6791 by including both crystallization and sedimentation.  Furthermore, it may be possible to separate these two effects by comparing observations of metal rich systems such as NGC 6791 to less metal rich systems such as NGC 6397 where sedimentation should be unimportant.  

Sedimentation depends on the diffusion constant $D$ for $^{22}$Ne ions in strongly coupled plasma mixtures.  There have been previous calculations of $D$, starting with the MD simulations of Hansen et al. for the one component plasma (OCP) \cite{Hanson75}.  The one component plasma consists of ions, with pure coulomb interactions, and an inert neutralizing background charge density.  Diffusion in the OCP in a strong magnetic field was considered by Bernu \cite{Bernu}.  Hansen et al. have also calculated diffusion for binary mixtures \cite{Hanson85}.           

Diffusion for a Yukawa fluid has been simulated by Robbins et al. \cite{Robbins} and Ohta et al. \cite{Ohta}.  In a Yukawa fluid ions interact via a screened coulomb potential $v_{ij}(r)$,
\begin{equation}
v_{ij}(r)=\frac{Z_iZ_j e^2}{r} {\rm e}^{-r/\lambda},
\label{v(r)}
\end{equation}
for two ions with charges $Z_i$ and $Z_j$, that are separated by a distance $r$. The OCP is equivalent to a Yukawa fluid, where all of the ions have the same charge and the screening length $\lambda$ is very large.

The motion of carbon, oxygen, and neon ions in a WD is largely classical because of their large masses.  However, at great densities there could be some quantum corrections that might increase $D$.  These have been estimated by Daligault and Murillo \cite{Daligault}, in a semiclassical model, and found to be very small.  Instead, these authors  argue that accurate classical simulations of diffusion coefficients in strongly coupled mixtures with impurities would be useful for WD cooling.  These diffusion coefficients would also be useful to describe sedimentation in neutron stars, see for example \cite{peng}. 

In this paper, we present classical MD simulations of carbon, oxygen, and neon plasma mixtures in order to determine diffusion coefficients $D$.   Our results are more accurate than previous work because we explicitly simulate mixtures with realistic WD compositions.  In Section \ref{sec.formalism} we describe our MD formalism and present results for diffusion coefficents in Section \ref{sec.results}.  We conclude in Section \ref{sec.conclusion}.

\section{Formalism}
\label{sec.formalism}

We describe our MD simulation formalism.  This is similar to what we used earlier to simulate the carbon/ oxygen phase diagram in ref. \cite{WD_PRL}.  We consider a three component mixture of carbon ($^{12}$C), oxygen ($^{16}$O), and neon ($^{22}$Ne), were the neon is assumed to have a small concentration.  A star with near solar metallicity, that has most of its original carbon, nitrogen, and oxygen converted into $^{22}$Ne, might have of order 2\% $^{22}$Ne.  The ratio of carbon to oxygen in the WD core depends on the rates for the $^4$He($2\alpha,\gamma$)$^{12}$C and $^{12}$C($\alpha,\gamma$)$^{16}$O reactions and is expected to be near one to one, see for example \cite{WD_PRL,wdevolution}.  Therefore in this paper we present simulations using a mixture of 49\% $^{12}$C, 49\% $^{16}$O, and 2\% $^{22}$Ne, by number.  We expect our results for the Ne self-diffusion coefficient to be nearly independent of $^{22}$Ne concentration, as long as it is small.  Likewise we do not expect much dependence of $D_{\rm Ne}$ on the ratio of carbon to oxygen.  In addition to this three component mixture, we also present results for a one component system of pure oxygen for comparison.

The ions are assumed to interact via screened Yukawa interactions, see Eq. \ref{v(r)}.  The Thomas Fermi screening length $\lambda$, for cold relativistic electrons, is $\lambda^{-1}=2\alpha^{1/2}k_F/\pi^{1/2}$ where the electron Fermi momentum $k_F$ is $k_F=(3\pi^2n_e)^{1/3}$ and $\alpha$ is the fine structure constant.  The electron density $n_e$ is equal to the ion charge density, $n_e=\langle Z\rangle n$, where $n$ is the ion density and $\langle Z\rangle$ is the average charge.  Our simulations are classical and we have neglected the electron mass (extreme relativistic limit).   This is to be consistent with our previous work on neutron stars.  However, the electron mass is important at the lower densities in WD and this may change our results slightly \cite{pot1}.     Also quantum effects could play some role at high densities \cite{pot2},\cite{jones}.  For relativistic electrons, the ratio of $\lambda$ to the ion sphere radius $a$,
\begin{equation}
a=\Bigl(\frac{3}{4\pi n}\Bigr)^{1/3},
\end{equation}
depends only on the average charge $\langle Z \rangle$.  For our three component mixture we use $\lambda=2.816 a$, while for pure oxygen, we use $\lambda=2.703 a$.  For nonrelativistic electrons $\lambda/a$ can be somewhat smaller.  In Sec. \ref{sec.results} we find that our results for diffusion constants $D$ are insensitive to a fifty percent decrease in $\lambda$.

The simulations can be characterized by an average coulomb parameter $\Gamma$,
\begin{equation}
\Gamma= \frac{\langle Z^{5/3} \rangle e^2}{a_e T}\, .
\label{gammamix}
\end{equation} 
Here $\langle Z^{5/3} \rangle$ is an average over the ion charges, $T$ is the temperature, and the electron sphere radius $a_e$ is $a_e=(3/4\pi n_e)^{1/3}$ with $n_e=\langle Z\rangle n$ the electron density.  The pure system freezes near $\Gamma=175$ \cite{pot1} while the C/O/Ne mixture is expected to freeze at a somewhat higher $\Gamma$ \cite{WD_PRL,CONe_phasediag}.  Note that these values of $\Gamma$ may depend slightly on $\lambda$ \cite{hamaguchi,pot1}.
 
Time can be measured in our simulations in units of one over the plasma frequency $\omega_p$.  Long wavelength fluctuations in the charge density can undergo oscillations at the plasma frequency.  This depends on the ion charge $Z$ and mass $M$.  For mixtures we define a hydrodynamical plasma frequency $\bar\omega_p$ from the simple averages of $Z$ and $M$,
\begin{equation}
\bar\omega_p=\Bigl[\frac{4\pi e^2\langle Z\rangle^2 n}{\langle M \rangle}\Bigr]^{1/2}.
\label{omega}
\end{equation}
Note that other choices for the average over composition in Eq. \ref{omega} are possible.  However, they are expected to give very similar results for the average plasma frequency.

Self diffusion constants $D_i$, for ions of type $i$, are calculated from the velocity autocorrelation function $Z^i(t)$,
\begin{equation}
Z^i(t)=\frac{\langle {\bf v}_j(t_0+t)\cdot {\bf v}_j(t_0) \rangle} {\langle {\bf v}_j(t_0)\cdot {\bf v}_j(t_0) \rangle}
\label{Z(t)}
\end{equation}
where the average is over all ions $j$ of a given type and over initial times $t_0$.  The velocity of the $j$th ion at time $t$ is ${\bf v}_j(t)$.  The diffusion constant is calculated from the time integral of $Z^i(t)$,
\begin{equation}
D_i=\frac{T}{M_i}\int_0^{t_{max}}dt Z^i(t).
\label{D_i(t)}
\end{equation} 
We find that $Z^i(t)$ is small by a maximum time of $t_{max}=220 /\bar\omega_p$.

We start our MD simulations from a random configuration at a relatively high temperature so that $\Gamma\approx 1$.  We evolve the system in time using the simple velocity Verlet algorithm \cite{verlet} with a time step $\Delta t=1/18\bar\omega_p$.  We use periodic boundary conditions.  We do not use a cutoff in the interaction at large distances and evaluate the force on a given ion by summing over all of the other ions in the simulation.  We test for finite size effects by performing simulations for $N=3456$, 8192, and 27648 ions.  Our largest simulation volume is large enough so that one half of the box length $L$ is much larger than the electron screening length $\lambda$, $L/2=8.65\lambda$.   We approximately maintain the system at a given temperature by simply rescaling the velocities every 10 time steps.  Typically we equilibrate the system by evolving for a time $2200/\bar\omega_p$.  Then we take data for an additional time of $2200/\bar\omega_p$ by writing the velocities to disk every $1/9\bar\omega_p$ (two time steps).  Next we rescale the velocities to a lower temperature and repeat the equilibration and data taking steps for the next higher value of $\Gamma$.  We present results for $Z^i(t)$ and $D_i$ in Section \ref{sec.results}  

\section{Results}
\label{sec.results}
\begin{figure}[ht]
\begin{center}
\includegraphics[width=3.5in,angle=0,clip=true] {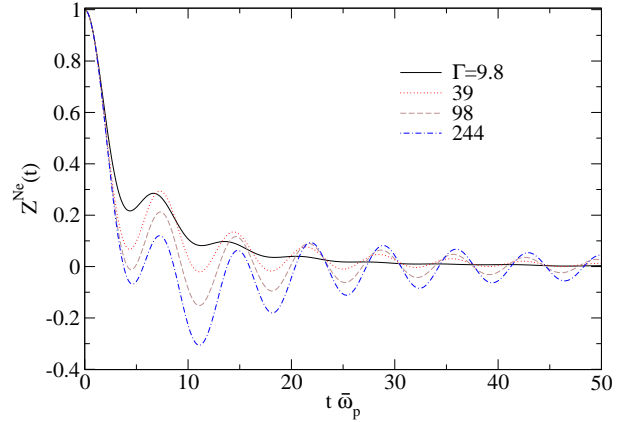}
\caption{(Color on line) Velocity autocorrelation function $Z^i(t)$ versus time $t$ for $^{22}$Ne ions in a carbon, oxygen, and neon mixture.  The N=8192 ion system is in a fluid state and the curves are for values of the coulomb parameter $\Gamma=$9.8 (solid), 39 (dotted), 98 (dashed), and 244 (dot-dashed). }
\label{Fig1}
\end{center}
\end{figure}

We now present results for the velocity autocorrelation function $Z^i(t)$ and diffusion constant $D_i$, first for a mixture of carbon, oxygen, and neon and then for a pure oxygen system.  Fig. \ref{Fig1} shows the velocity autocorrelation function for $^{22}$Ne from a simulation with $N=8192$ ions.  The velocity autocorrelation function oscillates with frequency near $\bar\omega_p$ and the amplitude of these oscillations increases with $\Gamma$.  Note that at $\Gamma=244$ the system is still in a (possibly metastable) fluid state.

\begin{figure}[ht]
\begin{center}
\includegraphics[width=3.5in,angle=0,clip=true] {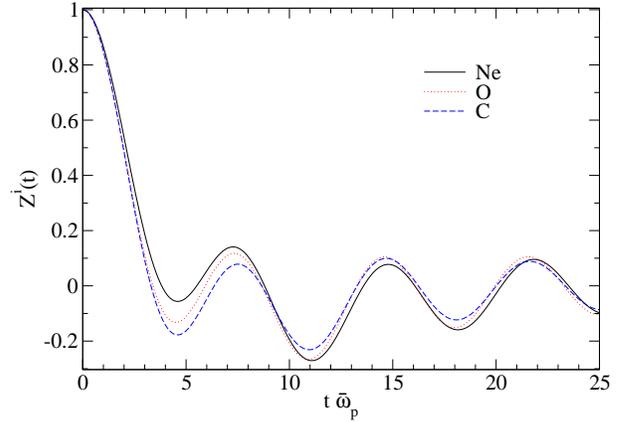}
\caption{(Color on line) Velocity autocorrelation function $Z^i(t)$ versus time $t$ for $i=^{22}$Ne (solid cuve), $^{16}$O (dotted), and $^{12}$C (dashed) ions.  The simulation has $N=8192$ ions at coulomb parameter $\Gamma=195.2$ and is in a liquid state. }
\label{Fig2}
\end{center}
\end{figure}

In Fig. \ref{Fig2} we show $Z^i(t)$ for $i=^{12}$C, $^{16}$O, and $^{22}$Ne ions.  The difference in $Z^i$ for different ions is subtle.  We see that $Z^i(t)$ at $t$ near $5/\bar\omega_p$ is more negative for $^{12}$C than for $^{22}$Ne.  Perhaps the lighter carbon ions bounce backwards from the confining cage of other ions with a more negative velocity than do neon ions.  These subtle differences in Fig. \ref{Fig2}, along with the explicit factor of $1/M_i$ in Eq. \ref{D_i(t)}, lead to differences in diffusion constants for different ions.

\begin{figure}[ht]
\begin{center}
\includegraphics[width=3.5in,angle=0,clip=true] {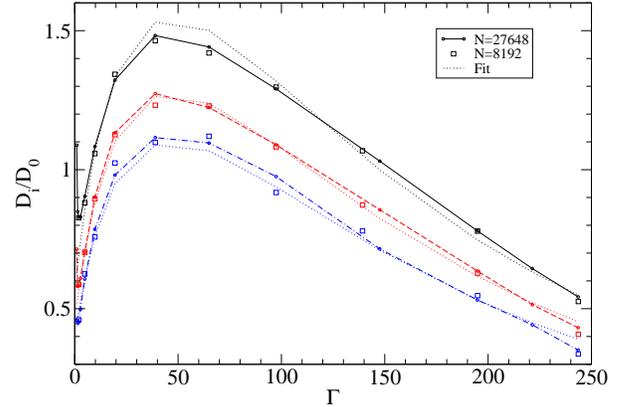}
\caption{(Color on line) Diffusion constants $D_i$ over a simple fit to the original MD results for the one component plasma $D_0$, see Eq. \ref{D_0}, versus coulomb parameter $\Gamma$.  Results for $^{12}$C (solid black line), $^{16}$O (dashed red), and $^{22}$Ne (dot-dashed blue) are presented for a simulation with 27648 ions.  Results for 8192 ions (squares) are also shown.  Finally the dotted lines show the simple global fit of Eq. \ref{fit}.}
\label{Fig3}
\end{center}
\end{figure}

We choose to scale our diffusion results with Hansen et al's simple fit to their original MD results for the diffusion constant $D_0$ of a one component plasma \cite{Hanson75}.  
\begin{equation}
D_0=\frac{3\bar\omega_pa^2}{\Gamma^{4/3}}
\label{D_0}
\end{equation}
Note that we have generalized ref. \cite{Hanson75} results to a multicomponent system by using the average plasma frequency $\bar\omega_p$ in Eq. \ref{D_0}.  In Table \ref{tableone} and Fig. \ref{Fig3} we present our MD simulation results for $D_i/D_0$.  We see that Eq. \ref{D_0} is indeed a reasonable first approximation to $D_i$ and our results only differ from this by about 50\% or less.  Finite size effects appear to be modest with 3456 ion simulation results being below 8192 or 27648 ion results by only a few \%.  In general there is good agreement between 8192 and 27648 ion results.  However, our results for $^{22}$Ne have larger statistical errors than for $^{12}$C or $^{16}$O because there are fewer $^{22}$Ne ions in our simulations, since its abundance is assumed to be only 2\%.  We test the sensitivity of our results to the screening length $\lambda$ by running a 27648 ion C, O, Ne system with a fifty percent smaller $\lambda$.   We find at $\Gamma=9.75$ that diffusion constants $D_i$ only increase by two percent or less.

\begin{table}
\caption{Diffusion constants $D_i/D_0$, for i=$^{12}$C, $^{16}$O, and $^{22}$Ne from simulations with $N=3456$, 8192 and 27648 ions, for different coulomb parameter $\Gamma$ values.  The diffusion constants are scaled with $D_0=3\bar\omega_p a^2/\Gamma^{4/3}$ which is a simple fit to the original one component plasma results, see Eq. \ref{D_0} in the text.}
\begin{tabular}{lllll}
$\Gamma$ & $N$& $^{12}$C & $^{16}$O & $^{22}$Ne \\
\toprule
0.974 &\ 3456 & & & \\
 &\ 8192  & & &\\
 &27648\ \  &1.092 & 0.717 & 0.460\\
 1.392 &\ 3456 & & & \\
  &\ 8192 & & & \\
   & 27648 & 0.852 & 0.582 & 0.449 \\
  1.948 &\ 3456 &0.827 &0.580 &0.477 \\
  &\ 8192 & 0.827& 0.587&0.460 \\
  & 27648 & 0.832 & 0.586 & 0.453 \\
2.783 &\ 3456 & & & \\
&\ 8192 & & & \\
& 27648 & 0.833 & 0.611 & 0.500 \\
4.871 &\ 3456 & 0.894 &0.697 & 0.572\\
&\ 8192&0.881 & 0.703 & 0.625 \\
& 27648 & 0.907 & 0.706 & 0.609 \\
9.742 &\ 3456 & 1.081 & 0.888& 0.749\\
&\ 8192 & 1.058 & 0.895 & 0.758 \\
& 27648 & 1.087 & 0.904 & 0.788 \\
 19.48 &\ 3456 & 1.311 &1.121 & 0.995\\
 &\ 8192 &1.343 & 1.126 & 1.024 \\
 & 27648 & 1.327 & 1.137 & 0.984\\
38.97 &\ 3456 & 1.455 &1.251 & 1.114\\
&\ 8192 & 1.464 & 1.232 & 1.098\\
&27648 & 1.488 & 1.278 & 1.119 \\
64.95 & \ 3456 & 1.446 &1.216 &1.076 \\
&\ 8192 & 1.420 & 1.229 & 1.120\\
&27648& 1.446&1.228 &1.100\\
97.42&\ 3456&1.294 & 1.077& 0.940\\
&\ 8192&1.297& 1.082&0.918\\
&27648&1.295&1.093 &0.979\\
147.6 &\ 3456 & & &\\
 &\ 8192 & & &\\
 &27648 & 1.034& 0.859 & 0.718\\
 194.84 & \ 3456 & & &\\
 &\ 8192 & 0.779 & 0.628 & 0.547\\
 &27648 & 0.783 & 0.638 & 0.533 \\
221.40 &\ 3456 & & &\\
&\ 8192 & & &\\
&27648& 0.646 & 0.516 & 0.443\\
243.55 & \ 3456 & & &\\
&\ 8192 & 0.526 & 0.408 & 0.337 \\
& 27648 & 0.544 & 0.433 & 0.352 \\ 

\end{tabular} 
\label{tableone}
\end{table}

The dependence of $D_i$ on ion species is very interesting.  The $^{12}$C, $^{16}$O, and $^{22}$Ne results in Fig. \ref{Fig3} are nearly parallel, for $\Gamma>$ few, and depend on the ion charge $Z_i$ approximately as $Z_i^{-2/3}$.  In general, $D_i$ depends on both the mass and the charge of the ion.  However, since the ions in our simulations have similar charge to mass ratios we do not separate out these two effects and the observed $Z_i^{-2/3}$ dependance includes both effects.  We can approximately fit all of the results in Fig. \ref{Fig3} with $\Gamma>5$ with a simple expression,
\begin{equation}
\frac{D_i}{D_0}\approx 0.53\Bigl[\frac{\langle Z\rangle}{Z_i}\Bigr]^{\frac{2}{3}} (1+0.22\Gamma)\exp(-0.135 \Gamma^{0.62}).
\label{fit}
\end{equation}   
This Eq. is shown as dotted lines in Fig. \ref{Fig3}.  

\begin{figure}[ht]
\begin{center}
\includegraphics[width=3.5in,angle=0,clip=true] {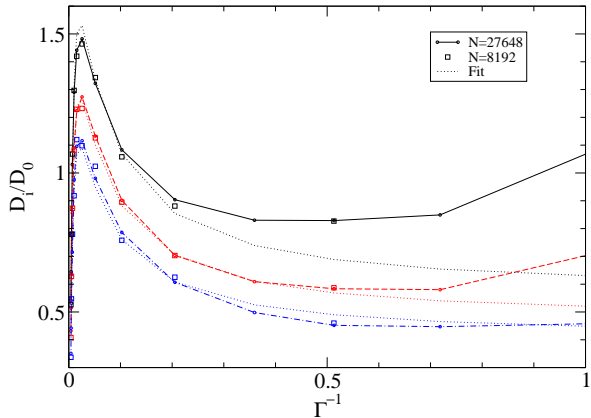}
\caption{(Color on line) Diffusion constants $D_i$ over $D_0$, see Eq. \ref{D_0}, versus the inverse of the coulomb parameter $1/\Gamma$.  Results for $^{12}$C (solid black line), $^{16}$O (dashed red), and $^{22}$Ne (dot-dashed blue) are presented for a simulation with 27648 ions.  Results for 8192 ions (squares) are also shown.  Finally the dotted lines show the simple global fit of Eq. \ref{fit}.  This disagrees with $^{12}$C and $^{16}$O data for small $\Gamma$}
\label{Fig4}
\end{center}
\end{figure}

For weak coupling (small $\Gamma$) the dependance of $D_i$ on $Z_i$ is stronger.  In the limit of very weak coupling, $D_i$ may be related to a mean free path which depends on one over an interaction cross section.  In Born approximation this cross section scales as $Z_i^2$ and gives a $D_i$ that depends more strongly on $Z_i$ than $Z_i^{-2/3}$.  We illustrate this by plotting the same data of Fig. \ref{Fig3} but as a function of $1/\Gamma$ instead of $\Gamma$ in order to show the low $\Gamma$ data more clearly.  We see that $D_i/D_0$ increases for $^{12}$C for small $\Gamma$ in a way that is not described by Eq. \ref{fit}.  Indeed Fig. \ref{Fig4} shows that $D_i$ depends more strongly on $Z_i$ than Eq. \ref{fit} for small $\Gamma$.

\begin{figure}[ht]
\begin{center}
\includegraphics[width=3.5in,angle=0,clip=true] {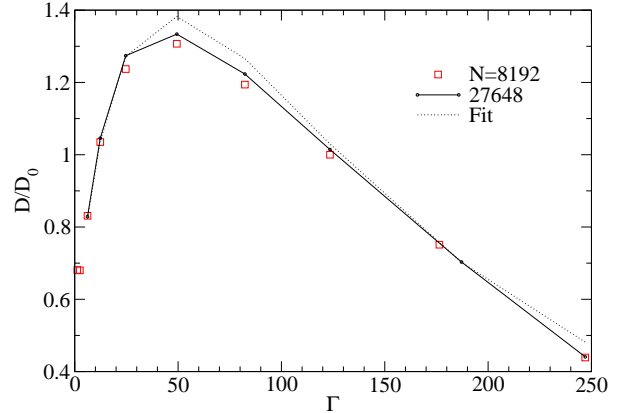}
\caption{(Color on line) Diffusion constant $D_i$ over $D_0$, see Eq. \ref{D_0}, versus the coulomb parameter $\Gamma$ for pure oxygen.  Results for a simulation with 27648 ions are shown as circles connected by a solid black line while the red squares show 8192 ion results.  Finally the dashed line shows the simple global fit of Eq. \ref{fit} with $\langle Z\rangle/Z_i=1$.}
\label{Fig5}
\end{center}
\end{figure}

Finally we also perform MD simulations for a one component system of pure $^{16}$O in order to compare to our multicomponent results.  Results for MD simulations with 8192 and 27648 ions are shown in Table \ref{tabletwo} and Fig. \ref{Fig5}.  There is good agreement between these two simulations and reasonable agreement with Eq. \ref{fit} with $\langle Z \rangle/Z_i=1$.  Previous MD simulations for one component Yukawa fluids fit $D$ with \cite{Daligault,Robbins,Ranganathan}
\begin{equation}
\frac{D}{\omega_pa^2}=0.0028 + 0.00525\Bigl(\frac{173}{\Gamma}-1\Bigr)^{1.154}.
\label{D}
\end{equation}  
The ratio of Eq. \ref{D} to Eq. \ref{D_0} is very similar in form to Fig. \ref{Fig5}, see Fig. 5 of Ref. \cite{Daligault}.  However, we find a somewhat larger amplitude for the deviations of $D$ from $D_0$.  

\begin{table}
\caption{Diffusion constant $D_{^{16}O}/D_0$, for a pure $^{16}$O system from simulations with $N=8192$ and 27648 ions in a (possibly metastable) fluid state.  The diffusion constant is scaled with $D_0=3\bar\omega_pa^2/\Gamma^{4/3}$.  This is a simple fit to the original one component plasma results, see Eq. \ref{D_0} in the text.}
\begin{tabular}{lll}
& $D/D_0$ & $D/D_0$ \\
$\Gamma$ & $N=8192$&N=27648  \\
\toprule
2.470& 0.680 & \\
6.175 &0.831 &0.828\\
12.35&1.035 & 1.045 \\
24.70&1.237 & 1.274\\
49.40& 1.307& 1.330\\
82.33 &1.194 & 1.223\\
123.50 &1.000 &1.014\\
187.12& & 0.703\\
247.00&.439 & 0.441\\
\end{tabular} 
\label{tabletwo}
\end{table}

\section{Conclusions}
\label{sec.conclusion}

Sedimentation of the neutron rich isotope $^{22}$Ne may be an important source of gravitational energy during the cooling of white dwarf stars.  This depends on the diffusion constant for $^{22}$Ne in strongly coupled plasma mixtures.  We have calculated self-diffusion constants $D_i$ from molecular dynamics simulations of carbon, oxygen, and neon mixtures.

We find that, for strong coupling (coulomb parameter $\Gamma>$ few), $D_i$ has a modest dependence on the charge $Z_i$ of the ion species, $D_i \propto Z_i^{-2/3}$.  However $D_i$ depends more strongly on $Z_i$ for weak coupling (smaller $\Gamma$).  Our results for both a carbon, oxygen, neon mixture, and for a one component plasma can be fit by,
\begin{equation}
\frac{D_i}{\bar\omega_p a^2}\approx 1.59 \Bigl(\frac{1+0.22\Gamma}{\Gamma^{4/3}}\Bigr)\exp(-0.135\Gamma^{0.62})\Bigl(\frac{\langle Z\rangle}{Z_i}\Bigr)^{2/3},
\end{equation}
for $\Gamma>$ few, see Eqs \ref{fit},\ref{D_0}.  Here $\bar\omega_p$ is the average plasma frequency and $a$ is the ion sphere radius.  We conclude that the self-diffusion constant $D_{Ne}$ for $^{22}$Ne in carbon, oxygen plasma mixtures is accurately known so that uncertainties in $D_{Ne}$ should be unimportant for simulations of white dwarf cooling.

All of our diffusion results have been for the liquid phase.  We are not aware of any results for $D_i$ in solids.  Often $D_i$ is arbitrarily assumed to be zero for a solid.  In future work we will simulate $D_i$ in the solid phase, both for single component and multicomponent systems.

We thank E. Brown, A. Cumming, and Z. Medin for helpful discussions.  This research was supported in part by DOE grant DE-FG02-87ER40365 and by the National Science Foundation through TeraGrid resources provided by National Institute for Computational Sciences, and Texas Advanced Computing Center under grant TG-AST100014.

\vfill\eject

\end{document}